\documentclass[prl,showpacs,twocolumn]{revtex4}

\usepackage{graphicx}
\usepackage{amsmath}
\usepackage{slashed}
\usepackage{feynmp}

\begin{document}
\title{London model of dual color superconductor}
\author{Ji\v r\'{\i} Ho\v sek}
\email{hosek@ujf.cas.cz} \affiliation{Department of Theoretical
Physics, Nuclear Physics Institute, Czech Academy of Sciences, 25068
\v Re\v z (Prague), Czech Republic\\
and\\
Institute of Experimental and Applied Physics, Czech Technical University in Prague,\\
Husova 240/5, 110 00 Prague 1, Czech Republic}

\begin{abstract}
\noindent Following closely the logic of the London phenomenological
macroscopic theory of the Meissner effect in superconductors we
describe the origin of the short-range behavior of the chromo-electric field, the necessary ingredient for color confinement in QCD.
The genuinely non-Abelian model is specified by the strong-coupling colored-gluon
current. Its first term, as the superconductivity current, is proportional to the gauge potential. The new term is simply related to
the chromo-magnetic pseudo-vector current of the non-Abelian Bianchi identity.
We suggest that this London dual color superconductivity current is responsible
for the observed almost perfect fluidity in droplets of the strongly interacting quark-gluon plasma.
Its chromo-magnetic component should have a specific experimental manifestation.
\end{abstract}

\maketitle

{\it Introduction.} Our trust in QCD relies mainly on its distinguished {\it
perturbative} property of asymptotic freedom \cite{wilczek}. It
uniquely describes the experimental facts in hadron physics caused
by the short-distance interactions of the colored quarks and the
colored gluons \cite{bjorken}. Another experimental fact, also
expected to be described by QCD, the permanent confinement of the
colored quarks and the colored gluons inside the colorless hadronic
jails, stays unexplained \cite{bjorken}. Yet it is being repeatedly confirmed by the extensive lattice QCD computations.
The color confinement is generally attributed
to the peculiar properties of the {\it non-perturbative} QCD vacuum
medium. The intuitive bag-model picture \cite{weisskopf} is that the
non-perturbative QCD vacuum is a medium which does not allow for the
penetration of the chromo-electric field, of which the colored
quarks and the colored gluons are the sources: Its chromo-dielectric
function $\epsilon$ vanishes.

As in the Lorentz-invariant theories the vacuum must look the same
in all Lorentz frames i.e.,
\begin{equation} \epsilon . \mu = 1,
\label{epsilon}
\end{equation}

\noindent the confining QCD vacuum behaves simultaneously as a
perfect color paramagnet: its chromo-magnetic permeability should
diverge: $\mu \rightarrow \infty$. Realizations of this general idea
started with \cite{models}. Particularly attractive are those which
employ the physical analogy with the "dual" phenomenon, the
confinement of the magnetic field in superconductors
\cite{superconductors}.

In perturbation theory the relation (\ref{epsilon}) is well
understood \cite{nielsen-hughes}: The asymptotic freedom or
anti-screening of the perturbative QCD vacuum medium ($\epsilon <
1$) results from a loop computation, and it is difficult to
understand in physical terms. The complementary relation $\mu > 1$
results from the summation over the Landau levels of the
color-charged particles in an external chromo-magnetic field, and it
is easy to understand: a weak color paramagnetism of the
perturbative QCD vacuum is due to the fact that the color-charged
QCD gluons carry the spin one \cite{johnson}.

The knowledge of the perturbative QCD Lagrangian is ultimately
sufficient for the computation of both the short- and the long-distance
quark and gluon properties. The lattice computations support
such a point of view. It is, however, quite conceivable
\cite{pagels-tomboulis} that for the intuitive, physical understanding of the
QCD quantum ground state, unlike in the infrared-free theories, the
classical QCD Lagrangian is practically useless. An effective QCD
Lagrangian seems much more efficient. We believe that in this
respect the lessons from the long-lasting development of
understanding superconductivity are highly relevant, and we try to
develop here the simplest one.

The phenomenon of superconductivity of electrons in metals at
temperatures close to the absolute zero was discovered in 1911 by
Camerling-Ones. In 1933 Meissner had found that the superconductors
are not only the perfect conductors but also the perfect diamagnets.
In 1935 H. London and F. London \cite{hflondon} came to the idea to
replace the Ohm's electric current $\vec j = \sigma \vec E$ in
Maxwell equations
\begin{equation}
{\rm div} \vec E = 0 \label{div}
\end{equation}
\begin{equation}
{\rm rot} \vec B - \frac{\partial}{\partial t} \vec E = \vec j
\label{rot}
\end{equation}
for the electric $\vec E$ and magnetic $\vec B$
fields by a super-current form $\vec j_s$. The reason is clear: For superconductors the
conductivity $\sigma$ in the Ohm's current is infinite. They
suggested for the superconducting current $\vec j_s$ the equations
\begin{equation}
\frac{\partial}{\partial t}\vec j_s = + \kappa \vec E \label{1}
\end{equation}
\begin{equation}
{\rm rot} \vec j_s = - \kappa \vec B \label{2}
\end{equation}

It is utmost important that these famous London equations are in
accord with the Bianchi identities (second pair of the Maxwell equations)
\begin{eqnarray}
{\rm div} \vec B \equiv 0\\
{\rm rot} \vec E + \frac{\partial}{\partial t} \vec B \equiv 0
\end{eqnarray}
which follow from the definition of the gauge-invariant field tensor
$F^{\mu \nu}$.

The equation (\ref{1}) is responsible for superconductivity
\cite{london}. The equation (\ref{2}) is responsible for the Meissner effect:
If we apply $\rm rot$ to (\ref{rot})
we get (for $\frac{\partial}{\partial t}\vec E = 0$) the equation
\begin{equation}
(\nabla^2 - \kappa)\vec B = 0
\end{equation}
which says that the magnetic field penetrates
into the body of superconductor only up to the London penetration
length $\lambda = 1/\surd \kappa$.

In the book from 1950 \cite{london} its author justifies the
equations (\ref{1}) and (\ref{2}) with a great charm: "The equations
at which we have arrived are distinguished by simplicity and
symmetry in such a way that we could hardly avoid writing them
down." F. London continues: "However, it is quite
conceivable that future developments may necessitate substantial
modifications of this theory".

Indeed, the modifications did follow, and it is fascinating that
the prescient Fritz London envisaged all of them \cite{london}: The first mandatory question
was: Where the equation
\begin{equation*}
\vec j_s = -\kappa \vec A
\end{equation*}
which clearly underlies the London equations might come from? Being not gauge-invariant Londons never used this equation
for the description of the properties of superconductors. F. London, however,
ingeniously speculated \cite{london} that such a current could be obtained from the {\it non-relativistic quantum-mechanical current} of many electrons
if {\bf superconductivity is a quantum phenomenon on macroscopic scale} characterized by
a kind of solidification or 'condensation of the average momentum distribution'.

Ginzburg and Landau (GL) realized that it could be the
electric current of the charged bosons {\it provided the boson field can
be replaced by a condensate}, and developed in 1950 their famous
phenomenological theory of superconductivity \cite{gl}. We notice
that both in the London and the GL theories the very notion of the
fermionic electrons as the carriers of the electric super-current is
entirely missing.

Finally, the second mandatory question was: How the (doubly charged)
scalar order parameter of Ginzburg and Landau can emerge in an
interacting many-electron system? The paradigm-changing answer was
found by Bardeen, Cooper and Schrieffer (BCS) in 1957 \cite{bcs}: It
is a Hartree-Fock-type mean field in the momentum-condensed ground state
of a many-electron quantum system with spontaneously broken fermion number.

Why the crude BCS approximation is so successful numerically was,
however, a mystery for years. It was clarified only in 1992 by
Polchinski \cite{polchinski}.

In the following we present what we think is the London theory of
the Meissner effect for the chromo-electric field. It requires
specification of the non-perturbative form of the color-gluon
current, and we briefly discuss its natural physical consequences.\\

{\it The QCD London equations.} We attempt here to follow the logic of the London theory, and
develop the analogous phenomenological description of the
experimental fact that the chromo-electric field becomes effectively
short-range at strong coupling. The equations of motion
corresponding to the perturbative QCD Lagrangian $L_{QCD}=
-\tfrac{1}{4}F^{\mu \nu}_a F_{a \mu \nu}$ (QCD Maxwell equations)
have the form
\begin{equation}
{\rm div} \vec E_a = -g f_{abc} \vec A_b.\vec E_c  \equiv \rho_a
\label{rho}
\end{equation}
\begin{equation}
{\rm rot} \vec B_a - \frac{\partial}{\partial t} \vec E_a =
-gf_{abc} \vec A_b \times \vec B_c \equiv \vec j_a \label{j}
\end{equation}

Here the chromo-electric field $\vec E_a$ and the
chromo-magnetic field $\vec B_a$ are defined as particular
components of the covariant color gluon field tensor
\begin{equation}
F^{\mu \nu}_a =\partial^{\mu}A^{\nu}_a-\partial^{\nu}A^{\mu}_a - g
f_{abc}A^{\mu}_b A^{\nu}_c.
\label{F}
\end{equation}

In the gauge $A^0_a=0$ (which, however, is not free of
subtleties \cite{bjorken}) they are defined as
\begin{equation}
\vec E_a = - \frac{\partial}{\partial t} \vec A_a \label{Ea}
\end{equation}

\begin{equation}
\vec B_a = {\rm rot} \vec A_a + \tfrac{1}{2}g f_{abc} \vec A_b
\times \vec A_c \label{Ba}
\end{equation}

Knowing that all analogies somehow falter we point out the distinct
properties of QCD:

1. In QCD the colored quarks seem entirely irrelevant for the
ultimate structure of the confining QCD vacuum medium, and are not
considered here. For practical applications, e.g. in hadron spectroscopy
the quarks are, however, indispensable as the sources of the
chromo-electric field. Their particular colorless and spin
configurations ultimately determine the shapes, spins and masses of hadronic bags.

2. The QCD Maxwell equations are non-linear, and cannot be expressed
entirely in terms of the chromo-electric and chromo-magnetic fields.
Both their left- and right-hand sides are gauge-dependent.

3. The Bianchi identities which follow from the definitions of $\vec E_a$
and $\vec B_a$ (second pair of the QCD Maxwell equations) have the
non-trivial right-hand sides:
\begin{equation}
{\rm div} \vec B_a = -g f_{abc} \vec A_b {\rm rot} \vec A_c \equiv
\rho^{(M)}_a \label{rhoM}
\end{equation}

\begin{equation}
{\rm rot} \vec E_a + \frac{\partial}{\partial t} \vec B_a = -g
f_{abc} \vec E_b \times \vec A_c \equiv \vec j^{(M)}_a \label{jM}
\end{equation}

The pseudo-scalar chromo-magnetic density $\rho^{(M)}_a$ was a sign
for the existence of the chromo-magnetic monopoles \cite{polyakov}.
Their condensate, as an analog of the condensate of the Cooper pairs
in superconductors, is in the heart of the underlying theories of
the dual color superconductors as the models of color confinement
\cite{superconductors}.

It is our understanding  that the pseudo-vector chromo-magnetic current $\vec j^{(M)}_a$ is another, more general
indication of the relevance of non-trivial chromo-magnetic configurations in the
confining QCD vacuum medium. In any case its $\rm rot$ will be present in the time derivative of the QCD London current.

4. The non-relativistic non-perturbative phenomena of
superconductivity are well described by theories at weak coupling.
In sharp contrast, the non-perturbative phenomena in QCD take place
only at large distances where their description requires the
strong-coupling tools.

{\bf We assume that there is a gauge and an appropriate, yet unknown, manifestly gauge-invariant
effective QCD Lagrangian at strong coupling} which yields, upon appropriate condensation, the QCD Maxwell equations in the form
\begin{equation}
{\rm div} \vec E_a = 0
\label{divE}
\end{equation}
\begin{equation}
{\rm rot} \vec B_a - \frac{\partial}{\partial t} \vec E_a = \vec J_a
\label{rotB}
\end{equation}
The intention is to find the explicit form of $\vec J_a$ which will yield
the Meissner effect for the chromo-electric field. It is natural to
expect that the new current will be relevant also for the
description of other QCD phenomena at strong coupling. In
particular, we guess it should describe some sort of super-fluidity
of the strongly coupled (predominantly gluonic) QCD matter.

First we apply the operation ${\rm rot}$ to the Eq.(\ref{jM}), use
the QCD equations of motion (\ref{divE}) and (\ref{rotB}) modified for the strong coupling, and
get
\begin{equation}
(\frac{\partial^2}{\partial t^2} - \nabla^2)\vec E_a  = -
\frac{\partial}{\partial t} \vec J_a - g f_{abc}
{\rm rot} (\vec E_b \times \vec A_c). \label{Ea}
\end{equation}
This equation suggests to postulate the first QCD London equation in
the form
\begin{equation}
\frac{\partial}{\partial t} \vec J_a = - \mu^2 \vec E_a - g f_{abc} {\rm rot} (\vec E_b \times \vec A_c)
\label{dtJ}
\end{equation}
for it leads to the equation for the Meissner effect of the
chromo-electric field
\begin{equation}
[(\frac{\partial^2}{\partial t^2} - \nabla^2) - \mu^2]\vec E_a = 0
\label{mu}
\end{equation}
Here $\lambda = 1/\mu$ is the London penetration length of the
chromo-electric field. It should be of the order of the typical
hadron size.

Second, we apply ${\rm rot}$ to (\ref{dtJ}), and after a simple
manipulation we obtain the second QCD London equation
\begin{equation}
{\rm rot} \vec J_a = + \mu^2 \vec B_a - \tfrac{1}{2} g f_{abc}[\mu^2
- {\rm rot}\phantom{i}{\rm rot}](\vec A_b \times \vec A_c)
\label{rotJ}
\end{equation}

It is readily verified that the new formulas (\ref{dtJ}) and
(\ref{rotJ}) for the chromo-electric and chromo-magnetic fields,
respectively, identically fulfil the second pair of the QCD Maxwell
equations (\ref{rhoM}) and (\ref{jM}).

In order to have the parallel with the ordinary superconductivity complete we present the QCD London equations
also for the case of the static chromo-electric field ( $\frac{\partial}{\partial t}\vec E_a = 0$):
It is easy to check that the equation
\begin{equation}
\frac{\partial}{\partial t} \vec J_a = + \mu^2 \vec E_a - g f_{abc} {\rm rot} (\vec E_b \times \vec A_c)\\
=+ \mu^2 \vec E_a + {\rm rot} \vec j^{M}_a
\label{dtJst}
\end{equation}
implies the finite penetration length for the static chromo-electric field:
\begin{equation}
[\nabla^2 - \mu^2]\vec E_a = 0
\label{must}
\end{equation}
The second London equation easily follows:
\begin{equation}
{\rm rot} \vec J_a = - \mu^2 \vec B_a + \tfrac{1}{2} g f_{abc}[\mu^2
+ {\rm rot}\phantom{i}{\rm rot}](\vec A_b \times \vec A_c)
\label{rotJst}
\end{equation}
Within our assumptions the gauge dependence of the QCD London current seems unavoidable. With some hesitation we write
\begin{equation}
\vec J_a = - \mu^2 \vec A_a + \tfrac{1}{2} g f_{abc} {\rm rot} (\vec A_b \times \vec A_c)
\label{J}
\end{equation}
and leave a discussion of this rather suspicious concept to the concluding section.

It is perhaps worth of noticing that in this case the signs of the "Abelian" pieces of $\frac{\partial}{\partial t} \vec J_a$
and ${\rm rot} \vec J_a$ proportional to $\vec E_a$ and $\vec B_a$, respectively, are identical to their London counterparts.
We take the parallel with the arguments of F. London seriously and discuss also the flow properties of $\vec J_a$
having in mind the London QCD equations (\ref{dtJst}) and (\ref{rotJst}).\\

{\it The colored gluon super-fluid.} Clearly, the 'derivation' of the QCD London equations depends upon the strong assumptions formulated above which can be verified
only {\it a posteriori} i.e., once the yet unknown gauge-invariant GL-like theory is found.
Phenomenologically the primary outcome, the Meissner effect for the chromo-electric field, is welcome:
it is in accord with the successful MIT bag model.

It is then natural to ask whether, similarly to the ordinary
superconductivity, the QCD London current describes some super-flow
of the strongly interacting QCD matter. "We could hardly avoid" a
speculation that it might be responsible for the observed almost perfect
fluidity of the strongly interacting colored gluon plasma
\cite{perfectfluid} (with an innocent role of the colored quarks). To demonstrate this convincingly deserves
extra work. Here we restrict ourselves merely to two qualitative remarks which hopefully support such a speculation.

First, it is gratifying that the suggestion respects the limited value of analogies. It is related to the point 4 above: Both the
London electrodynamics of ordinary conductivity and of superconductivity dealing with flow and super-flow
are the {\it macroscopic} descriptions. In contrast, the perturbative QCD Maxwell equations (\ref{rho}) a (\ref{j}) are by
definition appropriate for the {\it short-distance color-gluon
phenomena}. Consequently, the current $\vec j_a$
can hardly be compared with a macroscopic Ohm-like flow. The QCD
Maxwell equations (\ref{divE}) and (\ref{rotB}) are, however,
the effective ones, intended for the description of the {\it
large-distance color-gluon phenomena}. Hence, the long-distance flow
property associated with the current $\vec J_a$ should have the
reasonable macroscopic physical sense. By assumption, the "new" definitions of
the chromo-electric and chromo-magnetic fields $\vec E_a$ and $\vec
B_a$ in terms of the London QCD currents (\ref{dtJst}) and
(\ref{rotJst}), respectively, represent the natural physical degrees
of freedom. It is of course utmost important that they fulfil the
universally valid identities (\ref{rhoM}) and (\ref{jM}).

Second, association of $\vec J_a$ with the macroscopic flow of a strongly coupled
color-gluon super-fluid definitely amounts to a non-trivial modification of the velocity field due to the fact that
the time derivative of the London current contains the ${\rm rot}$ of the genuinely non-Abelian pseudo-vector magnetic current $\vec j^{M}_a$.
At the London phenomenological level such a description incorporates {\bf unspecified} chromo-magnetic degrees of freedom in the strongly
interacting gluon plasma. It is encouraging that in more explicit particular realizations of such a 'magnetic scenario' the plasma contains, besides the colored gluons,
also the chromo-magnetic monopoles, dyons or the magnetic strings \cite{shuryak}. In the condensed form these topological objects are suggested for the explanation of
the color confinement. The reliability of such a picture is supported by the lattice QCD simulations \cite{zakharov}.\\

{\it Conclusion and outlook.} Within the assumptions formulated above the present London model provides a natural phenomenological basis
for the whole class of models of the confining QCD vacuum medium viewed as a dual color superconductor: (1) It yields the Meissner effect for the chromo-electric field,
the necessary ingredient for the hadronic bag formation. (2) By definition, the QCD vacuum medium, a perfect color paramagnet,
is full of non-perturbative chromo-magnetic configurations. This is phenomenologically incorporated in the fixed
chromo-magnetic admixture in the London QCD current.
(3) The model hopefully describes also a sort of super-fluidity of the strongly interacting quark-gluon droplets \cite{krishna}. It is an unintended,
though quite natural, fortunate bonus. (4) In the macroscopic flow of the strongly interacting quark-gluon plasma
the unique chromo-magnetic component of the London QCD gluon current
should have the specific experimental manifestation.

At the same time the present London-type approach suggests that some physical significance
of the gauge potentials for the observable large-distance phenomena in QCD is unavoidable. The implementation of the gauge invariance
in general can be a rather subtle issue:
(i) In the Abelian electrodynamics the general expectation is
that the physical quantities should be gauge invariant i.e., the functions of the gauge invariant fields $\vec E$ and $\vec B$.
Yet, there is the famous quantum loophole, the experimentally observed Aharonov-Bohm effect \cite{ab}. (ii) Even more to the point, the subtlety of the gauge invariance
of the London theory of superconductivity mentioned in the Introduction is nicely discussed in the footnote of \cite{gsz}.
(iii) Finally and most important, the new phenomena related to the non-zero mean value $\langle A_{a\mu}A_a^{\mu}\rangle$ are expected to emerge
in the non-Abelian gauge theory (QCD) at strong coupling \cite{gsz,gz}.
Gubarev and Zakharov in their insightful paper \cite{gz} argue that the topological defects responsible for confinement (like the chromo-magnetic monopoles or the magnetic strings)
should manifest themselves by the gauge-dependent vacuum condensate $\langle A_{a\mu}A_a^{\mu}\rangle$.
Below we briefly illustrate how such a condensate might justify the form of our London QCD current (\ref{J}).

To the best of our knowledge the generally accepted effective theory
of the strongly interacting QCD (gluon) matter does not exist yet.
It should, for example, provide the field-theoretic description of the phenomenological MIT bag model.
In the superconductivity language the formation of the bags created by the magnetic
monopole-anti-monopole pair in the Abelian GL superconductivity
field was elaborated by Ball and Caticha \cite{ball-caticha}.
We believe that the general ideas of F. London about macroscopic quantum nature of the future molecular theory of superconductivity
apply also in the case of the strongly interacting QCD matter.

Pagels and Tomboulis \cite{pagels-tomboulis} have suggested the
non-polynomial effective theory in terms of $F^{\mu \nu}_a$. It results in
the particular dia-electric picture of the QCD vacuum. Another option
mentioned by these authors is to consider the algebraically
independent polynomials constructed from the QCD field tensor
$F^{\mu \nu}_a$. There are nine of them for the gauge group $SU(2)$
\cite{roskies}. As far as we know, for the physically distinct
$SU(3)$ even their list is not known.

We believe that the starting point for such a program, following
Steven Weinberg \cite{weinberg}, should be in principle infinite series of gauge-invariant terms
formed by the gauge-covariant $F^{\mu \nu}_a$ and its covariant derivatives.

As an illustration consider $L_{eff} = -\tfrac{1}{4}F^{\mu \nu}_a F_{a \mu \nu} + ...$
It is important that the higher derivatives in an effective Lagrangian are harmless in the sense that they do not cause the Ostrogradsky instabilities \cite{donoghue}.
Quantized is only the mandatory lowest-order term which defines the QCD degrees of freedom in terms of which we want to describe the system;  the higher orders are treated as interactions.

It is suggestive, following Ginzburg and Landau to replace in the most rigid terms of $L_{eff}$ the products of the two gauge vector potentials $A_a^i A_b^j$
by the corresponding Gubarev-Zakharov (GZ) vacuum condensate (Hartree-Fock-like mean field) i.e., by a constant
\begin{equation}
\langle A_a^i  A_b^j \rangle = \Lambda^2 \delta^{ij} \delta_{ab}
\label{AA}.
\end{equation}
In ordinary GL (or better the Abelian Higgs) model this step corresponds to replacing in the kinetic term $(D^{\mu} \Phi)D_{\mu} \Phi$ of the gauge-invariant Higgs Lagrangian
the complex scalar field $\Phi$ by its gauge-dependent vacuum expectation value $\langle \Phi \rangle = v$. From the resulting London Lagrangian $L_{London} = - e^2v^2 \vec A \vec A$
we easily get the (London) current $\delta L_{London}/\delta \vec A = "\vec j_s = -\kappa \vec A"$ for $\kappa = 2e^2v^2$.

Relying upon the usefulness of $\langle A^{\mu}_a A_{a \mu}\rangle$ \cite{gz} we suggest to proceed in $L_{eff}$ analogously:

First, consider in the mandatory, ordinary QCD term of $L_{eff}$ only its non-derivative part. In the gauge $A^0_a = 0$ it reads as
\begin{equation*}
-\tfrac{1}{4}g^2 f_{abc}f_{ade}\delta^{ij} \delta^{kl} A^i_b A^j_dA^k_c A^l_e.
\end{equation*}
Performing in this term one HF contraction (\ref{AA})
in all possible positions we arrive at $L_{eff}^{(1)} = -3g^2 \Lambda^2 \vec A_a \Vec A_a$. It yields the first term $- \mu^2 \vec A_a$
of the London QCD current (\ref{J}) for $\mu^2 = 6g^2 \Lambda^2$. It is perhaps interesting to note that even it is formally identical with
the ordinary London superconductivity current, the first term of the London QCD current is of the entirely different, genuinely non-Abelian origin.

Second, the straightforward idea is to obtain the second term of the London QCD current (\ref{J}) by similar procedure from appropriate dim 6 terms
in which we keep one derivative of the gauge potentials. The non-derivative dim 6 terms give rise to a correction to the first term using two contractions.
Finding in such a way the term $\tfrac{1}{2} g f_{abc} {\rm rot} (\vec A_b \times \vec A_c)$ will be the first step in our future work.

For us the most flagrant use of a gauge-non-invariant vacuum condensate is the glorious BCS \cite{bcs}:
Immediately after its advent Yoichiro Nambu asked himself \cite{nambu1}:
How can one trust the BCS with its gauge non-invariant two-fermion condensate
\begin{equation*}
\langle \Phi \rangle \sim \langle\psi_{\uparrow}\psi_{\downarrow}\rangle
\end{equation*}
for discussing the electromagnetic properties of superconductors
like the Meissner effect if its only excitations are the Bogoliubov \cite{bogo}- Valatin \cite{vala} fermionic quasi-particles not carrying the definite
electric charge ? He ingeniously saved the electric charge conservation and the gauge invariance of BCS
{\it by adding to it} \cite{nambu2} the now famous collective exitations, the Nambu-Goldstone bosons.

It is conceivable that the use of the GZ gauge-non-invariant but color-symmetry-conserving (!) condensate
together with appropriate account of the fixed topological defects (monopoles, strings, ..., ?) responsible for confinement will result
in a novel implementation of gauge invariance. The London QCD current (\ref{J}) seems indispensable for fixing the form
of an intuitively transparent effective strong-coupling QCD.

I am grateful to Tom Brauner, Petr Bene\v s and Adam Smetana for their interest in this work, and for valuable suggestions.

\end{document}